\documentclass{JHEP3}
\usepackage{epsfig}
\usepackage{graphicx}
\usepackage{dcolumn}
\usepackage{bm}
\newcommand{\be}{\begin{equation}}
\newcommand{\ee}{\end{equation}}
\def\bea{\begin{eqnarray}}
\def\eea{\end{eqnarray}}
\def\Tr{\mbox{Tr}}

\def\ket#1{| #1 \rangle}

\def\del {\partial}
\def\nn{\nonumber}
\def\aprime {\alpha^{\prime}}
\def\half {{1 \over 2}}

\def\parmedskipn        {  \par\medskip\noindent  }

 \def\be{\begin{equation}}
\def\ee{\end{equation}}
\def\bea{\begin{eqnarray}}
\def\eea{\end{eqnarray}}

\def\lesssim{\mathrel{\hbox{\rlap{\hbox{\lower4pt\hbox{$\sim$}}}\hbox{$<$}}}}
\def\gtrsim{\mathrel{\hbox{\rlap{\hbox{\lower4pt\hbox{$\sim$}}}\hbox{$>$}}}}

\title{String Theory and Hybrid Inflation/Acceleration}

\author{Carlos Herdeiro, Shinji Hirano and Renata Kallosh\footnote{Email addresses:\ carlos@het5.stanford.edu, hirano@itp.stanford.edu, kallosh@stanford.edu}\\
    Department of Physics, Stanford University, Stanford, CA 94305,
USA}
\received{\today}       
 \preprint{SU-ITP-01/44\\  ~hep-th/0110271\\ October 29, 2001}

\abstract{We find a description of hybrid inflation in
(3+1)-dimensions using brane dynamics of Hanany-Witten type.
P-term inflation/acceleration of the universe with the hybrid
potential has a slow-roll de Sitter stage and a waterfall stage
which leads towards an $\mathcal{N}=2$ supersymmetric ground
state. We identify the slow-roll stage of inflation with a
non-supersymmetric `Coulomb phase' with Fayet-Iliopoulos term.
This stage ends when the mass squared of one of the scalars in the
hypermultiplet becomes negative. At that moment the brane system
starts undergoing a phase transition via tachyon condensation to a
fully Higgsed supersymmetric vacuum which is the absolute ground
state of P-term inflation.  A string theory/cosmology dictionary
is provided, which leads to constraints on parameters of the brane
construction from cosmological experiments. We display a splitting
of mass levels reminiscent of the Zeeman effect due to spontaneous
supersymmetry breaking.}

\begin{document}


 \section{Introduction}

Recently it was pointed out in \cite{Kallosh:2001tm} that the
Salam-Strathdee-Fayet $\mathcal{N}=2$ SUSY gauge model
\cite{Salam:1975wa} provides a new type of inflationary model. The
theory has a $U(1)$ vector multiplet,  a charged hypermultiplet
and a Fayet-Iliopoulos (FI)  term.  In hypersymmetry ($\mathcal{N}=2$
supersymmetry) one has a triplet of prepotentials, $P^r$. They may
have some constant values $P^r= \xi^r$ that correspond to FI
terms in $\mathcal{N}$=2 supersymmetry. The cosmological theory based on this model was called in
\cite{Kallosh:2001tm} `hybrid hypersymmetric  model,' or P-term inflation, since the scalar
potential of this model corresponds to a hybrid-type potential \cite{Linde:1991km} with P-term.

P-term inflation is related to   D-term inflation theory \cite{Binetruy:1996xj},
 for the case when the gauge coupling $g$ is related to the Yukawa coupling $\lambda$ by  $
\lambda^2=  2 g^2 $. The potential also coincides with the F-term
inflationary potential studied in
\cite{Dvali:1994ms,Linde:1997sj}. Such
models are considered to be semi-realistic models of inflation in
the early universe (see for example
\cite{Linde:1997sj,Lyth:1997pf,Lyth:1999xn}). A nice introductory account
of the early universe acceleration (the cosmological inflation) and the present epoch acceleration can be found in \cite{Liddle:2000dt}.
In
\cite{Kallosh:2001tm} it has been suggested that P-term inflation
with  different parameters, gauge coupling  and FI terms, may also
be  used for explaining the  acceleration of the universe at the present epoch, with the
cosmological constant  $\Lambda = {\xi}^2/2$.

The purpose of this paper is to describe the connection between a
brane construction of string theory and the  cosmological aspects
of the hybrid hypersymmetric  model with P-term
inflation/acceleration. By making this connection we may constrain
parameters of the brane model using the recent cosmological
observations \cite{Smoot:1992td,supernova,CMB}.

Gauge theories related to brane configurations have been
extensively studied in the last few years, based on D-brane
technology motivated by the work of Polchinski
\cite{Polchinski:1995mt}-\cite{ Polchinski:1998rq}. The brane model
herein  is based on those of Witten \cite{Witten:1997sc} and
Hanany-Witten \cite{Hanany:1997ie}, which have been thoroughly
discussed and extended by Giveon and Kutasov \cite{Giveon:1999sr}.
Specifically, the model involves two parallel $NS5$-branes with a
$D4$-brane suspended between them and a $D6$-brane orthogonal to
both the $D4$-brane and the $NS5$-branes. Such system preserves
$\mathcal{N}=2$ $D=4$ supersymmetry. However,  a certain
displacement $\Delta L$ of the $NS5$-branes will break
supersymmetry spontaneously, as discussed recently by Brodie \cite{Brodie:2001pw}.
This provides positive vacuum energy which, from the
cosmological viewpoint, triggers inflation.

We will find that in our brane model one can express the FI term
 $ \xi$ as well as the gauge coupling $g$ of the cosmological model
 through a combination of the  string coupling $g_s$, string length $l_s$, the distance $L$ between heavy branes,  and their
 displacement $\Delta L$ from the supersymmetric position.\footnote{We use the
 conventional particle physics definition of the FI term with  $\Lambda=V= \xi^2/2$ and a conventional definition of gauge coupling, see
 \cite{Kallosh:2001tm} for more details. In cosmological applications \cite{Binetruy:1996xj,Lyth:1997pf,Lyth:1999xn} several  rescalings
 were made, both for the FI terms as well as for the charges. Therefore the expressions for the cosmological constant and $U(1)$ covariant
 derivatives in \cite{Binetruy:1996xj,Lyth:1997pf,Lyth:1999xn} are different from the canonical ones used here.}

 It has been shown
   \cite{Linde:1997sj,Lyth:1997pf,Lyth:1999xn} that for F and
 D-term inflation in the early universe with 60 e-foldings, some combination of the parameters of the relevant inflationary models can be defined  by the COBE
 measurement of CMB anisotropy \cite{Smoot:1992td} as follows: $\xi/ g  \sim  10^{-5} M_P^2$.\footnote{These relations are valid for $g^2\gtrsim 10^{-5}$ and
 ignoring the contribution of cosmic strings to perturbations of metric. For the relevant discussion see \cite{Kallosh:2001tm}.} In our brane model of P-term inflation  this yields the relation
 \bea \left( {\Delta L\over 2\pi g_s l_s} \right)_{\rm COBE}  \sim  10^{-5} \left({2\pi l_s\over
l_P}\right)^2\ . \eea
Here $ l_P^{-1}=M_P \sim 2.4 \times 10^{18}
\ GeV$.

Applying the model to the present epoch acceleration we can use
the indication from experiments on supernovae \cite{supernova} and
recent CMB observations \cite{CMB} that  the likely value of the
cosmological constant is $\Lambda_{present} \sim 10^{-120} M_P^4$.
Using the relation $\Lambda=\xi^2/2$, it follows that for the
brane model describing today's acceleration we get
 \bea
 \left({(\Delta L)^2\over 2 g_s l_s L}
 \right)_{\rm present} \sim  10^{-120}
 \left({2\pi l_s\over l_P}\right)^4 \ .
 \eea

We will point out the significance of the $\mathcal{N}=2$ FI terms
and show how the Coulomb branch, the mixed Coulomb-Higgs branch,
and the fully Higgsed branch  of the brane construction  are
related to the slow-roll de Sitter stage,  a waterfall stage, and
an absolute $\mathcal{N}=2$ supersymmetric ground state of the
hybrid hypersymmetric  model of inflation/acceleration in
\cite{Kallosh:2001tm}.\footnote{We use the term Coulomb branch
(Coulomb-Higgs branch) even for the case of time-dependent
non-vanishing vev's of the vector multiplet scalars and vanishing
vev's of the hypermultiplet scalars (scalars in both vector and
hypermultiplet are time dependent and non-vanishing). } The
connection is made more explicit by matching the masses of the
scalars in the hypermultiplet and the one-loop potential computed
from the field theory side with the ones computed from open string
theory.  In the absence of FI term neither our  gauge model nor
the brane construction lead to any interesting cosmological
models. However, when FI terms  are present, we find in the gauge
theory that the de Sitter type vacuum breaks supersymmetry
spontaneously, a fact which is also imprinted in the whole string
spectrum through the vanishing of the supertrace. Notice that a
spontaneously broken symmetry means that the underlying symmetry
 may still control the system as it happens for the standard model.

We will establish a relation  between Sen's tachyon condensation in open string theory \cite{Sen:1998sm} and tachyon condensation  in the context of  preheating in  hybrid inflation studied by Felder, Garcia-Bellido, Greene, Kofman, Linde and Tkachev \cite{FKL}. In P-term inflation,  when the system passes the  bifurcation point, the tachyonic  instability develops with the consequent waterfall to the $\mathcal{N}=2$ supersymmetric  ground state.

\section{The potential of the P-term inflation model}

The potential of the hybrid  hypersymmetric model
\cite{Kallosh:2001tm} is \bea V=  {g^2\over 2} \left[(|\Phi_1|^2 +
|\Phi_2|^2) |\Phi_3|^2 + |\Phi_1|^2 |\Phi_2|^2+ {1\over
4}\left(|\Phi_1|^2 -|\Phi_2|^2+ \frac{2{\xi}}{g} \right)^2\right]\
, \label{pot} \eea which is depicted in Figure 1.

\begin{figure}[h!]
\centering \epsfysize=9cm
\includegraphics[scale=0.7]{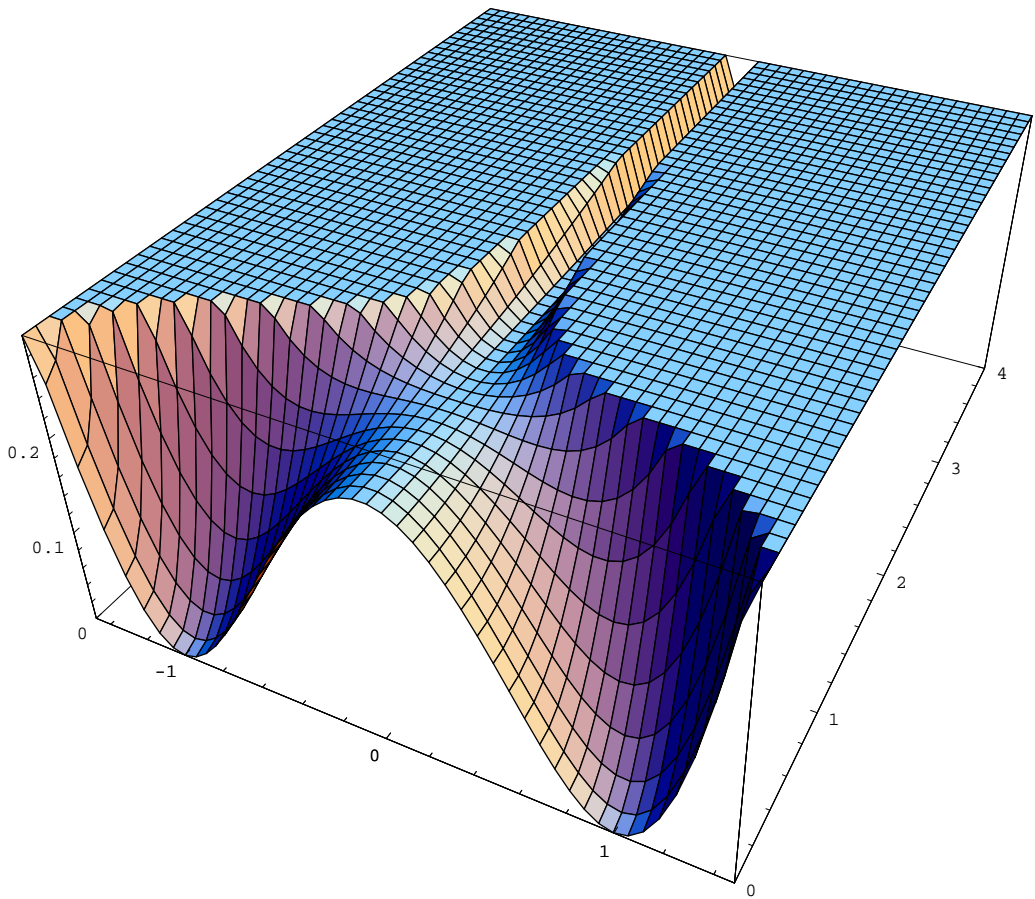}
\caption{Cosmological potential with Fayet-Iliopoulos term.
De Sitter valley is classically flat; it is lifted by the
one-loop correction corresponding to the one-loop potential
between $D4$-$D6$. In this figure the valley is along the
$|\Phi_3|$ axis; the orthogonal direction is a line passing
through the origin of the complex  $\Phi_2$ plane and we have put
$|\Phi_1|=0$. Notice there is no $\mathbb{Z}_2$ symmetry of the
ground state, it is just a cross section of the full $U(1)$
symmetry corresponding to the phase of the complex $\Phi_2$
field. The fields are shown in units of $\sqrt{\xi/ g}$. The bifurcation point corresponds to $|\Phi_3| =  \sqrt{\xi/g}$, $\Phi_2 = 0$. The absolute minimum is at $\Phi_3 = 0$, $\Phi_2=  \sqrt{2\xi/ g}$.} \label{}
\end{figure}

Here $\Phi_3\equiv  A+iB $ is a complex scalar from the
$\mathcal{N}=2$ vector multiplet and the two complex scalars
$\Phi_1 \equiv a_1+ib_1$ and  $\Phi_2 \equiv a_2+ib_2$ form a
quaternion of the  hypermultiplet, charged under the $U(1)$ group.
The FI P-term here is $\vec \xi=(0,0,\xi)$. All 6 real scalars
$\phi_i= \{A, B, a_1, b_1, a_2, b_2\}$ have canonically normalized
kinetic terms in the Lagrangian of the form \bea L_{\rm
kin}={1\over 2} \sum_{i=1}^{i=6}(\partial \phi_i)^2 = {1\over
2}|\partial \Phi_1|^2 + {1\over 2}|\partial \Phi_2|^2 + {1\over
2}|\partial \Phi_3|^2. \eea The potential (\ref{pot}) has a local
minimum, corresponding to a de Sitter space when coupled to
gravity, with $|\Phi_3|$ being a flat direction. These classical
vacua break all the supersymmetry spontaneously; here, the vev of
the hypers vanishes, $\langle \Phi_1 \rangle = \langle \Phi_2
\rangle =0$,  and the vev of the scalar from the vector multiplet,
which is the inflaton field, is non-vanishing, $\langle \Phi_3
\rangle \neq 0$. The masses of all the fields in the de Sitter
valley are as follows:  in the vector multiplet the gauge field
$A_\mu$ and a gaugino $\lambda_A$ are massless, whereas the masses
of the fields in the hypermultiplet are split: \bea
M^2_{2}=g^2|\Phi_3|^2 - g{\xi}\ , \qquad M_{\psi} = g |\Phi_3| \ ,
\qquad M_1^2=g^2|\Phi_3|^2 +g {\xi}\ . \label{split}\eea Here
$\psi$ is the hyperino, $\Phi_1$ ($\Phi_2$) are positively
(negatively) charged scalars of the hypermultiplet. The value of
the potential at this vacuum is $V= {\xi}^2/ 2$. This is the
cosmological constant driving the exponential expansion of the
universe. This state corresponds to a Coulomb branch of the
$\mathcal{N}=2$ gauge theory. The presence of the FI term breaks
supersymmetry spontaneously, which is imprinted in the fact that
the supertrace of the mass spectrum vanishes \cite{Ferrara:1979wa}
\begin{equation}
\mbox{STr}\, M^2\equiv \sum_{j}(-1)^{2j}(1+2j)M_j^2 =0,
\label{str}\end{equation}
where $j$ is the spin of the state. The right hand side of this equation
vanishes in our case since the total $U(1)$ charge vanishes for the hypermultiplet.

The point where one of the
scalars in the hypermultiplet becomes massless, \bea M^2_{2}=
g^2|\Phi_3|_c^2 - g {\xi}=0 \ \  \Leftrightarrow \ \  |\Phi_3|_c =
\sqrt {\xi\over g} \eea is a bifurcation point. At $|\Phi_3|^2\leq
{\xi}/g$, the de Sitter minimum becomes a de Sitter maximum;
beyond it, such scalars become tachyonic. The system is unstable
and the waterfall stage of the potential leads it to a ground
state. The waterfall stage has  non-vanishing vev's for the
scalars in both the hyper and  vector multiplets; this is a mixed
Coulomb-Higgs branch. Finally, the system gets to the absolute
minimum with vanishing vev for the scalars in the  vector
multiplet, $\langle \Phi_3 \rangle =0$, and non-vanishing vev for
the scalars in hypermultiplet,  $\langle \Phi_2 \rangle ^2=
2{\xi}/g$. Supersymmetry is unbroken and all fields are massive;
they form a massive $\mathcal{N}=2$ vector multiplet with
$M^2=2g{\xi}$. This is a fully Higgsed branch of the gauge theory.

Since the potential is flat in the $|\Phi_3|$ direction, the
inflaton field, $\Phi_3$, does not naturally move. However, the
gauge theory one-loop potential lifts the flat direction, via a
logarithmic correction
\cite{Dvali:1994ms,Binetruy:1996xj,Linde:1997sj,Lyth:1997pf,Lyth:1999xn}
\bea V = {1\over 2} {\xi}^2 + {g^2\over 16 \pi^2}
{\xi}^2 \ln { |\Phi_3|^2\over  |\Phi_3|_c^2}.
\label{1loop}\eea This is precisely what is necessary to provide a
slow roll-down for the inflaton field, since it is an attractive
potential which leads to the motion of the field $\Phi_3$ towards
the bifurcation point and to the end of inflation. Note that in
$\mathcal{N}=2$ supersymmetric gauge models there are no higher
loop infinities \cite{Grisaru:1982zh}.

Before proceeding to the string theory model for hybrid inflation,
we would like to stress that in the absence of FI term $\xi$ none
of the interesting things takes place. The potential with $\xi=0$
is plotted in Figure 2. There is a Minkowski valley with the flat
direction.
\begin{figure}[h!]
\centering \epsfysize=8cm
\includegraphics[scale=0.7]{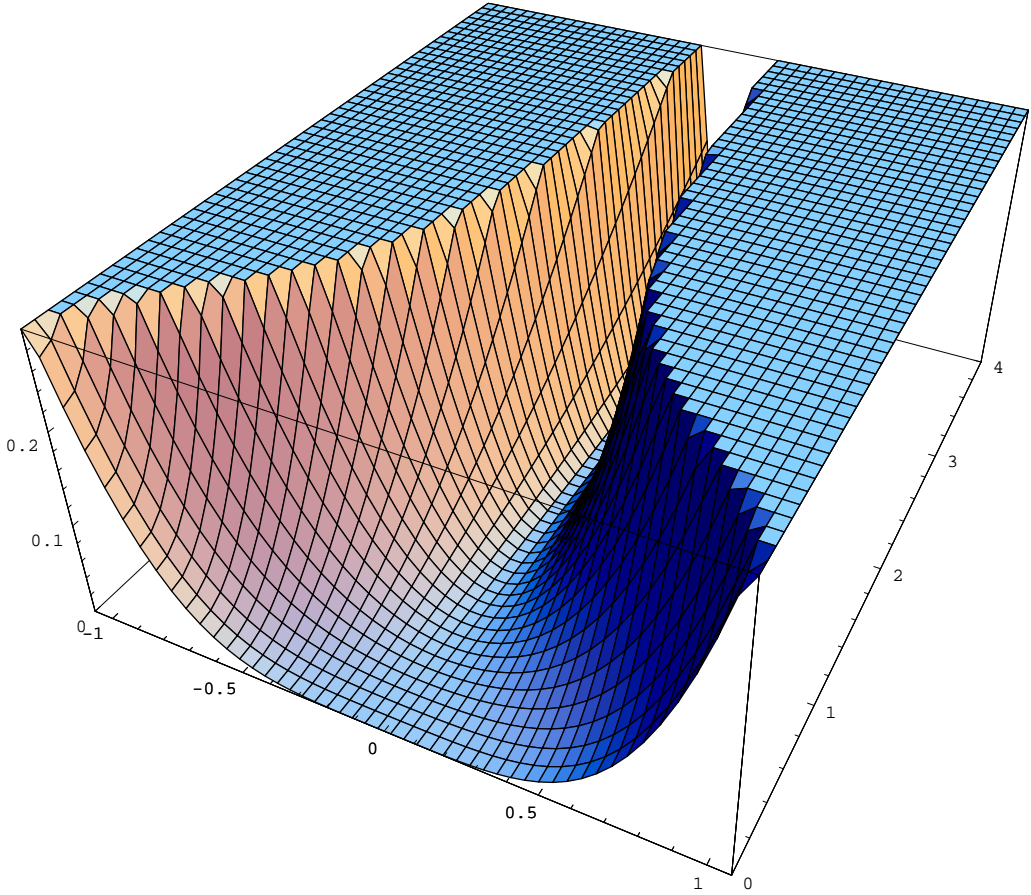}
\caption{Cosmological potential without Fayet-Iliopoulos term. The
motion of the D4 corresponds to moving along the bottom of the
valley, which has a zero potential.} \label{pot1}
\end{figure}

\section{ NS5-D4/D6-NS5 model}

\noindent We will now discuss one of the possible brane
constructions describing hybrid inflation. The picture is based on
the one studied in  \cite{Witten:1997sc}-\cite{Brodie:2001pw} in
(3+1)-dimensions. As a warm up, let us explain the supersymmetric
version of this construction, corresponding to the absence of FI
terms, which is depicted in Figure 3.

The following table summarizes the brane configuration, where $\times$s indicate  directions in which branes are lying.
\begin{center}
 \begin{tabular}{*{11}{|c}|}
 \hline
 \mbox{ } & 0 & 1 & 2 & 3 & 4 & 5 & 6 & 7 & 8 & 9 \\ \hline
 $D4$ & $\times$ & $\times$ & $\times$ & $\times$ & $\hspace{0.3cm}$ & $\hspace{0.3cm}$ & $\times$ & \mbox{} & \mbox{} &  \\ \hline
 $D6$ & $\times$ & $\times$ & $\times$ & $\times$ & \mbox{  } & \mbox{  } & \mbox{} & $\times$ & $\times$ & $\times$ \\
 \hline
 $NS5$ & $\times$ & $\times$ & $\times$ & $\times$ & $\times$ & $\times$ & $\mbox{} $ & \mbox{} & \mbox{} &  \\ \hline
 \end{tabular}
\end{center}
This brane model consists of two $NS5$-branes with a $D4$-brane
suspended between them. The field theory on the $D4$-brane is
effectively (3+1)-dimensional since one of the $D4$-brane
worldvolume directions is finite, with length $L$. Thus,
Kaluza-Klein modes may be ignored as long as we are probing
energies smaller than $1/L$, and physics is effectively
(3+1)-dimensional in the worldvolume theory. The $NS5$-branes play
still another role. They freeze the motion of the $D4$-brane in
the $7,8,9$ directions. Hence, such scalars will not appear in the
worldvolume theory; the only scalars arising therein correspond to
motions in the $4,5$ direction, and form the two real scalars of
the $\mathcal{N}=2$ vector multiplet.

\begin{figure}[h!]
\begin{picture}(0,0)(0,0)
\end{picture}
\centering 
\includegraphics[scale=0.8]{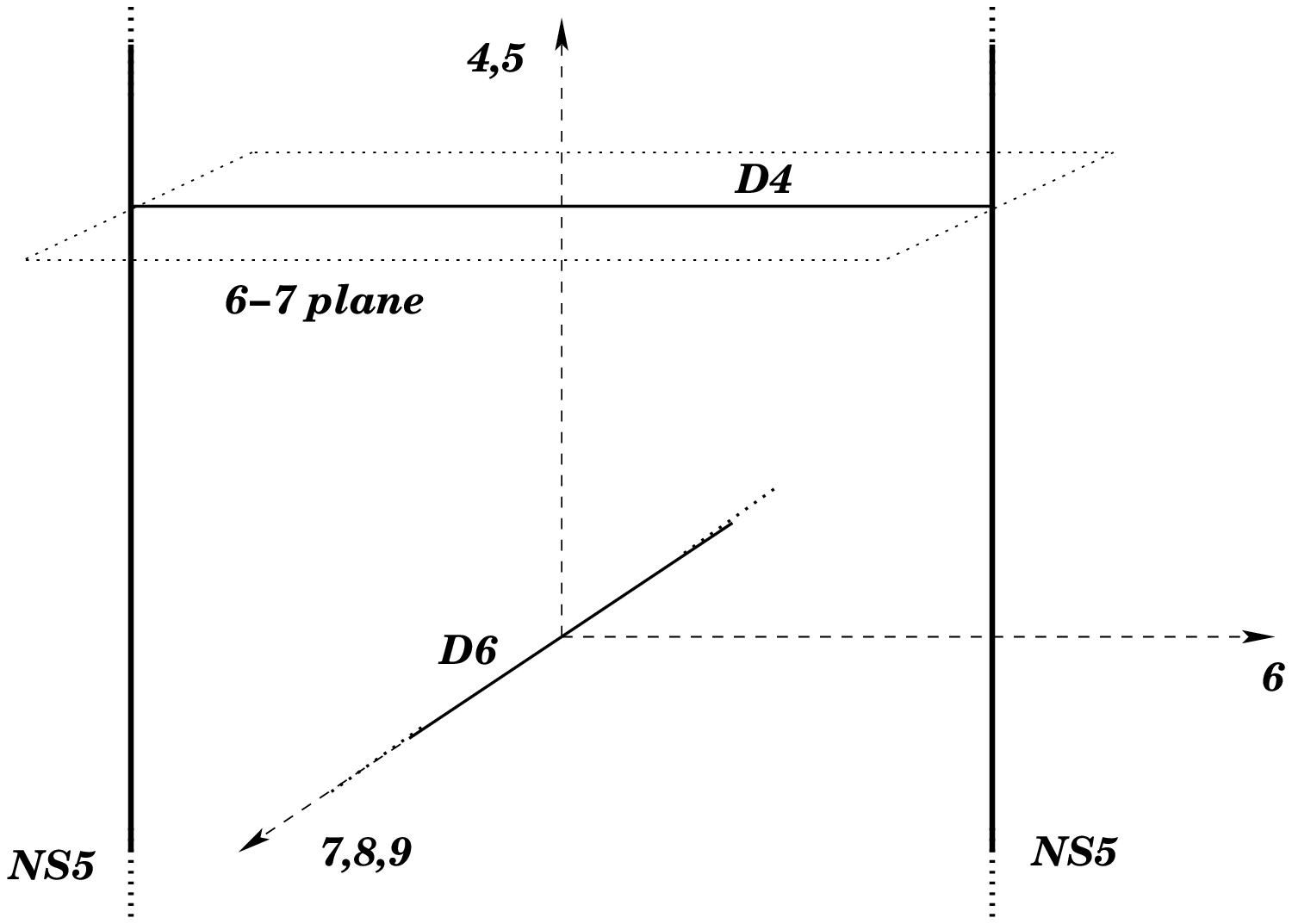}
\caption{Brane configuration without Fayet-Iliopoulos term. We are free to move D4 in 4,5 directions with no energy cost.}
\label{ctccre}
\end{figure}

In order to include matter in the $D4$-brane worldvolume theory,
we introduce a $D6$-brane. The (4-6) and (6-4) strings will then
form an $\mathcal{N}=2$ hypermultiplet. The spectrum of these
strings is given with some detail in Appendix A. Moreover,
including the $D6$-brane does not break any further supersymmetry,
since the projectors of the supersymmetry conditions are
compatible with the ones of the $NS5$ and $D4$. Therefore, the
worldvolume theory on the $D4$-brane is a (3+1)-dimensional,
$\mathcal{N}=2$  $U(1)$ gauge theory with one charged
hypermultiplet. Of course this is exactly the theory discussed in
section 2 without the FI term. Due to supersymmetry we are free to
move the $D4$-brane along the directions $4,5$ at no energy cost.
This corresponds to motion along the Minkowski valley in Figure 2
and the Coulomb branch of the gauge theory.

A much more interesting situation takes place when we turn on the
Fayet-Iliopoulos term, as emphasized in the last section for
cosmological applications. Consider displacing the $NS5$-branes
along direction 7 as shown in Figure 4a). Since the $D4$-brane has
to remain connected to the $NS5$-branes, this introduces an angle
$\phi$ between the $D4$-brane and the $D6$-brane, which in general
breaks supersymmetry. As we shall see, this angle $\phi$
corresponds in the field theory language to the FI parameter.

\begin{figure}[h!]
\centering 
\includegraphics[scale=0.55]{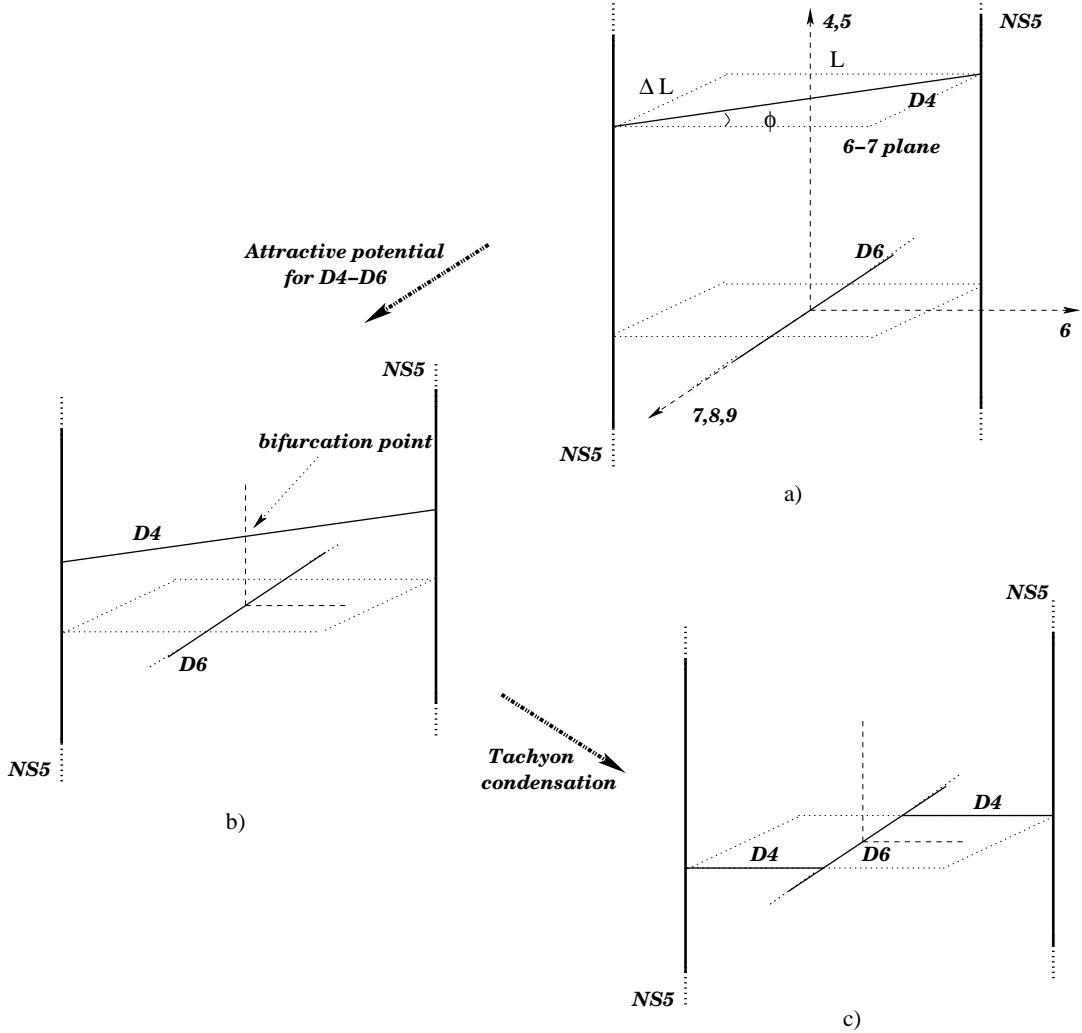}
\caption{Brane configuration evolution with Fayet-Iliopoulos term. a) For $\phi\neq 0$, supersymmetry is broken and D4-D6 experience an attractive force. b) At the bifurcation point, a complex scalar in the hypermultiplet becomes massless; when we overshoot  tachyon instability forms, taking the system to a zero energy ground state  shown in c) .}
\end{figure}

If initially $\phi \neq 0$, the system is unstable. One consequence is the attractive potential driving the $D4$-brane towards the
$D6$-brane, with the former sliding down the $NS5$-branes.
One might also think that the $D4$-brane tries to minimize the angle with the
$D6$-brane, by effectively trying to pull the $NS5$-branes to
the origin of direction 7. The latter effect causes the bending of the $NS5$-branes. We do not expect such an effect to drive the system towards a supersymmetric and hence minimal energy configuration, since the $NS5$-branes will be bent. In particular it cannot bring the configuration back to the supersymmetric system in Figure 3. Of course the effect can be always negligible if we consider large $NS5$-brane tension corresponding to weak string coupling.

The most important dynamical effect is therefore the attractive potential between $D4$ and $D6$. We will show below that the open string theory one-loop potential matches exactly (\ref{1loop}), which is derived from field theory. Therefore the motion of the $D4$ towards the $D6$ is the slow roll down of the inflaton. Hence it is a Coulomb branch with vanishing vev's for the scalars in the hypermultiplet and non-vanishing vev's for scalars in the vector multiplet.

To make the connection with the Coulomb phase more concrete at
this stage we use the spectrum of 6-4 and 4-6 open strings in the
presence of the angle $\phi$. We can see from Figure 4 that the
boundary conditions are slightly unusual in the 6-7 plane.  In
terms of $Z=X^6+i X^7$ and the rotation angle $\phi$ we require:

\begin{eqnarray}
\sigma_1&=&0\quad(D4)\qquad
\del_{\sigma_1}\rm{Re}\left(e^{i\phi}Z\right)=
\rm{Im}\left(e^{i\phi}Z\right)=0, \nonumber\\
\sigma_1&=&\pi\quad(D6)\qquad \del_{\sigma_1}\rm{Im} \ Z=\rm{Re} \
Z=0, \label{46bnd}
\end{eqnarray}
For $\phi=0$, these reduce to ordinary Dirichlet-Neumann (DN)
boundary conditions. The calculation of the spectrum of low lying
states of the open strings exactly reproduces the split in the
hypermultiplet masses shown in eq. (\ref{split}). By comparing
these two spectra we will identify the dictionary between the
parameters of the brane construction and cosmology.

\section{String theory--cosmology dictionary}

The relevant part of the string spectrum obtained in Appendix A
gives \bea M^2_{\pm}= {g_{{}_{YM}}^2(\widetilde{\Delta s})^2 \over
l_s^4} \pm {g_{{}_{YM}}\tilde{\phi}\over 2\pi l_s^2}\ , \qquad
M_{\psi}^2={g_{{}_{YM}}^2 (\widetilde{\Delta s})^2 \over  l_s^4}\
, \label{stringm}\eea \bea (\widetilde{\Delta s})^2\equiv
{(x^4)^2+(x^5)^2 \over g_{{}_{YM}}^2{\pi^2}}\ . \eea Notice that
despite the fact that the gauge theory is abelian, we use the
Yang-Mills subscript $g_{{}_{YM}}$, as a reminder that it is a
gauge coupling of the fields on the brane. We define the string
length as $\alpha' = l_s^2$. Since the kinetic terms on the field
theory side are canonically normalized, we had to redefine the
fields from the open string theory side such that
\begin{equation}
\Delta s= g_{{}_{YM}}(\widetilde{\Delta s}), \ \ \ \ \phi=
g_{{}_{YM}}\tilde{\phi}.
\end{equation}
For the Fayet-Iliopoulos parameter we use the convention of
\cite{Salam:1975wa}, for which the cosmological constant depends
solely on the FI parameter (not on the coupling). In this fashion
the fields in the $D$-brane action, which has initially in front a
$1/g_s$ factor, will get the same canonical normalization. Using
the conventions of Polchinski \cite{Polchinski:1996na} and taking
into account the compact dimension, the relation between couplings
is
\begin{equation}
g_{{}_{YM}}^2=(2\pi)^2g_s {l_s \over L}. \end{equation}

Comparison between formulae (\ref{split}) and (\ref{stringm})
yields the dictionary between field theory parameters (left) and
string theory parameters (right)
\begin{eqnarray}
g & \longleftrightarrow & g_{{}_{YM}}=2\pi \sqrt{g_s \frac{l_s}{L}}, \nonumber \\
|\Phi_3| & \longleftrightarrow & \frac{\widetilde{\Delta s}}{l_s^2}, \\
{\xi} & \longleftrightarrow & \frac{\tilde{\phi}}{2\pi
l_s^2} \nonumber.
\end{eqnarray}
For small angles $ L \phi=  L \sin \phi$, which is the distance in
direction 7 between the  NS5 branes after they have been pulled
out. We will therefore use the notation $L \phi \equiv \Delta L$.
The FI term is related to the string construction by a simple
formula which in string units simply states that it is a ratio
between the pull out distance and the finite size of the
$D4$-brane \bea {\xi}= {\Delta L \over L} {1\over 2\pi
l_s^2}{1 \over g_{{}_{YM}}}. \end{eqnarray}

As we have explained in the
beginning of the paper, a combination of the FI term $\xi$ and gauge coupling $g$ is constrained by recent
cosmological observations and it is very nice to find out a
possible interpretation of these important parameters in string
theory.

\section{Spontaneous supersymmetry breaking, potential  and tachyon condensation}

In field theory, spontaneous breaking of supersymmetry manifests
itself through the vanishing of the supertrace, as in equation
(\ref{str}). We have seen this to be the case in the field theory
description of the Coulomb phase, corresponding in the
cosmological picture to the slow-roll period of inflation. Of
course, the field theory contains only the low lying states of the
string theory. So it is natural to ask if the supertrace vanishes
for the whole tower of string states in the $D4$-$D6$ system  with
an angle. To check this, we only need the partition function,
$\mathcal{Z}=\mathcal{Z}(q)$, which is shown in appendix B. Then,
the supertrace can be expressed as
\begin{equation}
\mbox{STr}M^2=\left.
{2 \over \aprime}q{\del \over \del q}\mathcal{Z}
\right|_{q=1}.
\label{strace}
\end{equation}
An explicit calculation shows that the supertrace is indeed
vanishing. In fact, we find it rather impressive that there is a
mass splitting such that the supertrace vanishes at each level. We
illustrate such `stringy Zeeman effect' in figure 5. This
indicates that supersymmetry is spontaneously broken in the full
string theory.

\begin{figure}[h!]
\begin{picture}(75,0)(0,0)
\put(10,389){$\alpha' M^2=\frac{(\Delta s)^2}{\alpha '}$}
\put(-5,279){$\alpha' M^2=\frac{(\Delta s)^2}{\alpha
'}+\frac{1}{2}$} \put(-5,98){$\alpha' M^2=\frac{(\Delta
s)^2}{\alpha '}+1$}
\end{picture}
\centering \epsfysize=13cm
\includegraphics[scale=0.5]{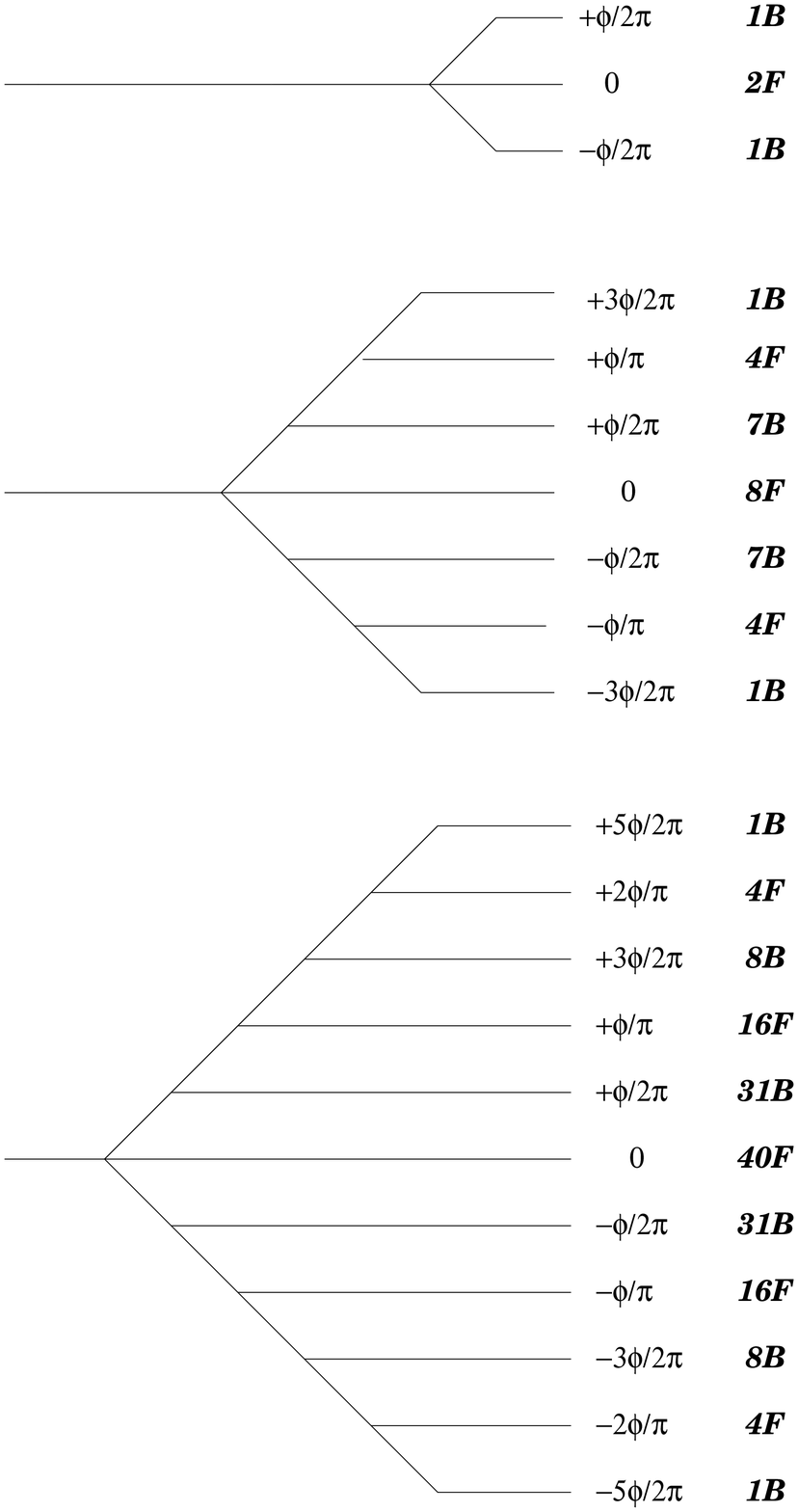}
\caption{Splitting of mass for the first three levels due to the
presence of an angle $\phi$. Notice that at each level the
supertrace vanishes. On the right we show the number of bosonic
(B) or fermionic (F) states with such mass for either the 4-6 or
6-4 strings. The total number of states is twice as many.}
\end{figure}

We may now proceed to  Figure 4b). In gauge theory, the one-loop
quantum correction (\ref{1loop}) to the classical potential,
drives the inflaton towards the bifurcation point at $|\Phi_3|^2=
\xi/ g$. The open string one-loop potential is expected to provide
an analogous attractive potential between the $D4$ and
$D6$-branes. The one-loop vacuum amplitude corresponding to the
effective interaction between $D4$ and $D6$-branes is given by
\begin{eqnarray}
V&=& -\frac{2}{2}\int_0^{\infty} {dt \over
t}\int_{-\infty}^{\infty} \frac{d^4p}{(2\pi)^4}\left[
\Tr_{NS}{1+(-1)^F \over 2}e^{-2\pi t L_{0}^{NS}}-\Tr_{R}{1+(-1)^F
\over 2}e^{-2\pi t L_{0}^{R}}\right].
\label{bpot}
\end{eqnarray}
The operator $\half(1+(-1)^F)$ is the GSO projection. The factor
of $2$ in the coefficient is due to the contribution of both 4-6
and 6-4 strings. Some details of the calculation can be found in
appendix C. The result is
\newpage
\begin{eqnarray}
V&=& \left({1 \over 8\pi^2\aprime}\right)^2 {\sin^2\phi \over
\cos\phi} \int_{\aprime/\Lambda^2}^{\infty} {dt \over t}
\exp\left[-2\pi t {(\Delta s)^2 \over \aprime} \right]
+{\cal O}(e^{-\pi/t}), \nn\\
 &\sim& {g_{{}_{YM}}^2 \over 16 \pi^2 }\left(\frac{\tilde{\phi}}{2\pi \aprime}\right)^2
\log{(\Delta s)^2  \over \Lambda^2}.
\label{stringpot}\end{eqnarray}
This exactly reproduces the one-loop correction in the field
theory (\ref{1loop}), including the numerical coefficient, in the
small angle and large separation approximation. Notice that the
logarithmic dependence is expected. In fact, the $D6$ and $D4$
branes have two common transverse directions. Since at large
separation the dominant contribution comes from the massless
closed string exchange, we expect to be dealing with a harmonic
potential in two-dimensions. Notice also that the logarithmic
divergence we have regularized with the cutoff $\Lambda$
corresponds, from the viewpoint of closed string theory, to an
infrared divergence. In the open string channel it is an
ultraviolet divergence and originates from the highly massive open
string modes, rather than the low lying states that contribute to
the one-loop field theory correction. Therefore, it is non-trivial
that we obtain a precise matching with the one-loop field theory
potential. Similar phenomena have been found in
\cite{Douglas:1997yp}, \cite{Banks:1997vh}.

The potential is attractive and drives the motion of the $D4$
brane towards the $D6$-brane. At a separation defined by the
vanishing of the mass of the lowest lying state \bea  (\Delta
s)^2_B= {\phi  \over 2\pi }l_s^2,\ \eea there is a bifurcation
point; beyond it, such state becomes tachyonic. Naturally, if the
inflaton field gets to the bifurcation point with non-zero
velocity, it will overshoot, and the tachyon instability will
develop. This is in precise correspondence with the motion on the
ridge of the potential in Figure 1 after the bifurcation point,
since the de Sitter valley becomes a hill top there, hence an
unstable maximum. The actual behaviour of perturbations in such
potentials has been investigated numerically in \cite{FKL} where
the tachyonic instability was studied in the context of preheating
of the universe after inflation.

Naively, one might expect the fields $\Phi_2$ and $\Phi_3$ to roll
down from the bifurcation point and then experience a long stage
of oscillations with amplitude $O(\sqrt{\xi / g})$ near the
minimum of the effective potential at $|\Phi_2| = \sqrt{2\xi / g}$
until they give their energy to particles produced during these
oscillations. However, it was recently found in \cite{FKL} that
the tachyonic instability rapidly converts most of the potential
energy $\xi^2/2$ into the energy of colliding classical waves of
the scalar fields. This conversion, which was called ``tachyonic
preheating,'' is so efficient that symmetry breaking (tachyon
condensation) is typically completed within a single oscillation
of the field distribution as it rolls towards the minimum of its
effective potential.

It is interesting to clarify the connection between such tachyonic
instabilities in models of hybrid cosmology and the phenomenon of
tachyon condensation in open string theory first discussed by Sen
\cite{Sen:1998sm}. Typically, the attention in the latter studies
was towards brane/antibrane systems with the consequent
brane/antibrane annihilation with the tachyon potential being
conjectured to cancel completely the brane's tensions (note that
an antibrane is a brane rotated by $\pi$ relative to the first
one; the supersymmetric configuration requires the same type of
branes to be parallel).

In fact, interesting brane inflation models have been suggested in
the framework of brane/antibrane configurations \cite{BAB}. In
this framework  tachyon instability develops at brane/antibrane
separations of the order of $l_s$, with the consequent tachyon
condensation. It is suggested therein that such inflation may be
of the hybrid inflation type, which is exactly coming out in our
study.

 The tachyonic instability of our system has also been
considered in the context of tachyon condensation in open string
field theory \cite{David:2000um}. Therein, the system considered
is $D0$-$D4$, which is just T-dual to our orthogonal $D4$-$D6$.
Moreover, instead of introducing an angle to create the
instability, a $B$ field is used. These are just T-dual pictures.
When the tachyon develops, it takes the system to a supersymmetric
ground state, which is a non-threshold bound state of $D0$ and
$D4$-branes. This mass deficit, is precisely the height of our de
Sitter valley in Figure 1. The perturbative string theory
techniques do not allow us to trace the evolution of the system
once the tachyonic instability develops, corresponding to the
waterfall stage in the cosmological picture. However, using the
open string field theory techniques of \cite{David:2000um} one
might be able to trace down quantitatively such evolution.

In our case, we know from the cosmological part of our
construction that after the bifurcation point the waterfall stage
(tachyon condensation) takes place. Instead of continuing at the
ridge (in Coulomb phase) the brane system undergoes a phase
transition. The system reconfigures itself as to reach the
supersymmetric configuration represented in Figure 4c). This is
the only supersymmetric configuration possible if we allow only
the $D4$-branes to move. It is an $\mathcal{N}=2$ supersymmetric,
fully Higgsed phase of the field theory. All fields are massive.
This absolute ground state is in precise agreement with the
Minkowski ground state of the hybrid potential in Figure 1. Notice
that this final configuration, where branes reconnect, is very
different from the brane/antibrane scenarios where the branes
annihilate after tachyon condensation.

\section{Discussion}
There are well known problems in incorporating de Sitter space and
cosmology into the framework of M/String theory; see the most
recent discussions in
\cite{Hull:2001ii,Spradlin:2001pw,Townsend:2001ea}. In this paper
we have described hybrid P-term inflation through a brane
construction of string theory. It suggests a path to link string
theory and cosmology. One may expect that eventually this
direction will be useful in addressing the  problems of cosmology
where  general relativity and gauge theory break down and quantum
gravity regime takes place. Here, we only gave a few steps towards
realizing a scenario of cosmological inflation/acceleration in
string theory. Including gravity is, of course, the main
challenge, since at present it is not known how to embed our
hybrid  hypersymmetric model into $\mathcal{N}=2$ supergravity;
only the coupling to $\mathcal{N}=1$ supergravity is known
\cite{Kallosh:2001tm}.

To include four dimensional gravity in our model we have to resort
to compactification, since no promising suggestions on how to
localize gravity on $D$-branes have been put forward. But string
compactifications of models with branes are very constrained.
Physically the flux of the branes in a compact space has to be
cancelled. A known example of a consistent string compactification
with $D5$ and $D9$-branes was worked out by Gimon and Polchinski
\cite{Gimon:1996rq}. Such models yield  $\mathcal{N}=1$ string
vacua in six dimensions, containing not only gravity but also the
vector and hyper multiplets of the field theory described in this
paper, which have a natural origin as $\mathcal{N}=1$ multiplets
in six dimensions.

We would like to emphasize again the important role of the
Fayet-Iliopoulos term in getting interesting cosmology. In our
model it has a simple geometric interpretation, as a displacement
of the $NS5$-branes. If we would pursue the suggestion of the
previous paragraph, we would have to understand how the FI term
arises there. It is known, that it has a geometric interpretation
as controlling the resolution of singularities in ALE spaces
\cite{Douglas:1996sw}. A compact version of such spaces is
$T^4/{\mathbb Z}_2$. This is an orbifold limit of $K3$, which is
precisely the compactification manifold used in
\cite{Gimon:1996rq}. In fact these two interpretations of the FI
terms are not unrelated, since it has been argued that under
certain circumstances the existence of $NS5$-branes is T-dual to
singularities of ALE spaces \cite{Brunner:1998gf}. It is amusing
to think that, in this picture, one could argue for the existence
of a positive cosmological constant by requiring smoothness of the
ten dimensional space of string theory. Thus, it seems important
to understand the coupling of our $\mathcal{N}=2$ gauge theory to
$\mathcal{N}=2$ supergravity  and to find its string theory
interpretation.

\section *{Acknowledgments}
We benefited from discussions with K. Dasgupta, E. Halyo, L. Kofman, D. Kutasov, A. Linde, A. Maroto, S. Shenker, E. Silverstein and L. Susskind.  This work is supported by NSF grant PHY-9870115. C.H is supported by grant SFRH/BPD/5544/2001 (Portugal). S.H was supported by the Japan Society for the Promotion of Science.

\

\appendix
\section {D4-D6 system with angle}
In order to address an audience of non string theory experts, we will give some details of the computation of the string spectrum in this appendix.

We consider the system of $D4$ and $D6$-brane at an angle and separated by some distance. The configuration is as in Figure 4a). We introduce a spacetime complex coordinate $Z=X^6+iX^7$. The boundary conditions of several types of open strings are summarized in the following table. We will denote Neumann and Dirichlet boundary conditions by N and D respectively.

\def\Re {\mbox{Re}}
\def\Im {\mbox{Im}}
\begin{center}
 \begin{tabular}{*{9}{|c}|}
 \hline
 $\mbox{}$ & $X^{0,1,2,3}$ & $X^{4,5}$ & $\Re Z$ & $\Im Z$ & $\Re (e^{i\phi}Z)$ & $\Im (e^{i\phi}Z)$ & $X^{8,9}$ \\ \hline
4-4 & NN & DD & - - & - - & NN & DD & DD \\ \hline
6-6 & NN & DD & DD & NN & - - & - - & NN \\ \hline
4-6 & NN & DD & -D & -N & N- & D- & DN \\ \hline
6-4 & NN & DD & D- & N- & -N & -D & ND \\ \hline
 \end{tabular}
\end{center}

We parametrize the open string worldsheet by $w=\sigma_1+i\sigma_2$ with $0\le\sigma_1\le\pi$ and $-\infty\le\sigma_2\le +\infty$. The presence of the angle $\phi$ is reflected in the spectrum of the 4-6 and 6-4 strings. The boundary conditions for 4-6 string in $(6,7)$-plane are somewhat unusual and are given in eq. (\ref{46bnd}),
while those in the rest of eight directions satisfy standard NN, DD or DN conditions. Given these boundary conditions, the solution to the equation of motion, $\del_w\del_{\bar{w}}X^{\mu}=0\quad (\mu=0,1,\cdots,9)$, yields the following expansion for 4-6 string.

\parmedskipn
\underline{($0,1,2,3$)-directions (ordinary NN)}
\begin{eqnarray}
X^i(w,\bar{w})=x^i-2i\aprime p^i\sigma_2
+i\sqrt{{\aprime \over 2}}\sum_{m\in Z\backslash\{0\}}
{\alpha_m^i \over m}\left(e^{imw}+e^{-im\bar{w}}\right),
\end{eqnarray}
\parmedskipn
\underline{($4,5$)-directions (ordinary DD)}
\begin{eqnarray}
X^{4,5}(w,\bar{w})={x^{4,5} \over \pi}\sigma_1
+i\sqrt{{\aprime \over 2}}\sum_{m\in Z\backslash\{0\}}
{\alpha_m^{4,5} \over m}\left(e^{imw}-e^{-im\bar{w}}\right),
\end{eqnarray}
\parmedskipn
\underline{($6,7$)-directions (rotated ND and DN)}
\begin{eqnarray}
Z(w,\bar{w})&=&i\sqrt{{\aprime \over 2}}e^{-i\phi}
\left[
\sum_{r\in Z+\half+{\phi \over \pi}}{1 \over r}
\alpha_re^{irw}+\sum_{r\in Z+\half-{\phi \over \pi}}
\bar{\alpha}_r e^{-im\bar{w}}\right],\\
\bar{Z}(w,\bar{w})&=&i\sqrt{{\aprime \over 2}}e^{+i\phi}
\left[
\sum_{r\in Z+\half-{\phi \over \pi}}{1 \over r}
\bar{\alpha}_re^{irw}+\sum_{r\in Z+\half+{\phi \over \pi}}
\alpha_r e^{-im\bar{w}}\right],
\end{eqnarray}
where $\alpha_r$ and $\bar{\alpha}_r$ are linearly independent, and their hermitian conjugates satisfy $\alpha^{\dagger}_r=-\alpha_r$ and $\bar{\alpha}^{\dagger}_r=-\bar{\alpha}_r$.

\parmedskipn
\underline{($8,9$)-directions (ordinary DD)}
\begin{eqnarray}
X^{8,9}(w,\bar{w})=
i\sqrt{{\aprime \over 2}}\sum_{m\in Z+\half}
{\alpha_m^{8,9} \over m}\left(e^{imw}-e^{-im\bar{w}}\right).
\end{eqnarray}
The mode expansions for fermions $\psi^{\mu}\,(\mu=0,1,\cdots,9)$ are similar. In the NS sector, the modes are shifted by $\half$ from those of bosons $X^{\mu}$, while they are the same in the R sector. It is convenient to introduce a complex fermion $\psi=\psi^6+i\psi^7$, as we have done for bosons. Then we have, for instance, in the NS sector,
\begin{eqnarray}
\psi(w,\bar{w})&=&e^{+i\phi}\left[
\sum_{r\in Z+{\phi \over \pi}}i^{-1/2}\psi_r e^{irw}
+\sum_{r\in Z-{\phi \over \pi}}
i^{1/2}\bar{\psi}_r e^{-ir\bar{w}}\right],\\
\bar{\psi}(w,\bar{w})&=&e^{-i\phi}\left[
\sum_{r\in Z-{\phi \over \pi}}i^{-1/2}\bar{\psi}_r e^{irw}
+\sum_{r\in Z+{\phi \over \pi}}
i^{1/2}\widetilde{\bar{\psi}}_r e^{-ir\bar{w}}\right],
\end{eqnarray}
and in the R sector, the modes are shifted by $\half$. Now the mass-shell condition is given by $L_0=0$ and one can easily find
\begin{equation}
L_0^{NS}=N_B+N_F^{NS}-\aprime M^2 +{(\Delta s)^2 \over \aprime} -{\phi
\over 2\pi},
\end{equation}
for the NS sector, and
\begin{equation}
L_0^{R}=N_B+N_F^R-\aprime M^2 +{(\Delta s)^2 \over \aprime},
\end{equation}
for the R sector. We have defined
\begin{equation}
(\Delta s)^2= {(x^4)^2+(x^5)^2 \over \pi^2}.
\end{equation}

It is now easy to read off the low lying states of open strings. Taking into account the GSO projection, the lowest mass state turns out to be the NS ground state $\ket{p^i;s_3=-1/2,s_4=-1/2}$ (where $p^i$ is 4-dimensional momentum) with mass
\begin{equation}
M^2_-={(\Delta s)^2 \over (\aprime)^2} -{\phi \over 2\pi\aprime},
\end{equation}
corresponding to two real scalars (together with 6-4 string) in the hypermultiplet. Then the next lowest state is built by $\bar{\psi}_{-\phi/\pi}(\psi^8_0+i\psi^9_0)\ket{p^i;s_3=-1/2,s_4=-1/2}=\ket{p^i;s_3=+1/2,s_4=+1/2}$ with mass
\begin{equation}
M^2_+={(\Delta s)^2 \over (\aprime)^2} +{\phi \over 2\pi\aprime},
\end{equation}
giving the remaining half of scalars in the hypermultiplet. The
spacetime fermions in the hypermultiplet turns out to be the R
ground state $\ket{p^i;s_1,s_2}$ with $s_1s_2=-1/4$ by the GSO
projection, whose mass is
\begin{equation}
M^2_{\psi}={(\Delta s)^2 \over (\aprime)^2}\ ,
\end{equation}
corresponding to two 4-dimensional Weyl spinors (together with 6-4 string).

It is straightforward to carry out the computation of the spectrum for 4-4 and 6-6 strings. So we will not repeat the analysis in these cases, but the massless spectra amount to the dimensional reductions of an ${\cal N}=1$ vector multiplet in 10-dimensions down to 5- and 7-dimensions respectively. In our case we are only interested in the gauge theory of the $D4$-brane suspended between two $NS5$-branes, so 6-6 strings decouple from the dynamics and also the collective excitations of the $D4$-brane given by 4-4 strings in 6,7,8 and 9 directions are frozen due to the suspension of the $D4$-brane between two $NS5$-branes. Thus the massless spectrum of 4-4 strings reduces to an ${\cal N}=2$ vector multiplet in 4-dimensions.

\section{ The Supertrace of $M^2$}
In order to compute the supertrace (\ref{str}) we need the partition function.
 The partition function for the bosonic part is evaluated as
\begin{eqnarray}
\mathcal{Z}_B=\Tr e^{-2\pi t\left(N_B+N_{bc}\right)}
&=&\prod_{m=1}^{\infty}(1-q^m)^{-4}
(1-q^{m-1/2-\phi/\pi})^{-1}
(1-q^{m-1/2+\phi/\pi})^{-1} \nn \\
&&\times
(1-q^{m-1/2})^{-2},
\end{eqnarray}
where we introduced $q=e^{-2\pi t}$. In this expression, $N_{bc}$ is the number operator for the ghosts. Similarly we will introduce  $N_{\beta \gamma}^{NS}$ and $N_{\beta \gamma}^{R}$. The partition function for the fermionic part is computed as
\begin{eqnarray}
\mathcal{Z}_{NS}=\Tr_{NS}{1+(-1)^F \over 2}
e^{-2\pi t\left(N_F^{NS}+N_{\beta\gamma}^{NS}\right)}
&=&\half\cdot 2\prod_{m=1}^{\infty}(1+q^{m-1/2})^{4}(1+q^m)^{2}\nn \\
&&\times
(1+q^{m-\phi/\pi})(1+q^{m-1+\phi/\pi})
\end{eqnarray}
in the NS sector, and
\begin{eqnarray}
\mathcal{Z}_R=\Tr_{R}{1+(-1)^F \over 2}
e^{-2\pi t\left(N_F^{R}+N_{\beta\gamma}^{R}\right)}
&=&\half\cdot 4\prod_{m=1}^{\infty}(1+q^m)^{4}(1+q^{m-1/2})^{2}\nn \\
&&\times
(1+q^{m-1/2-\phi/\pi})(1+q^{m-1/2+\phi/\pi})
\end{eqnarray}
in the R sector. The full partition function for either the 4-6 or
6-4 strings therefore amounts to
\begin{eqnarray}
\mathcal{Z}&=&q^{(\Delta s)^2/\aprime}
\prod_{m=1}^{\infty}(1-q^m)^{-4}(1-q^{m-1/2-\phi/\pi})^{-1}
(1-q^{m-1/2+\phi/\pi})^{-1}(1-q^{m-1/2})^{-2} \nn\\
&\times&
\left[q^{-\phi/(2\pi)}\prod_{m=1}^{\infty}
(1+q^{m-1/2})^4(1+q^{m-\phi/\pi})
(1+q^{m-1+\phi/\pi})(1+q^{m})^{2}\right. \nn\\
&&\left.-2\prod_{m=1}^{\infty}
(1+q^{m})^4(1+q^{m-1/2-\phi/\pi})
(1+q^{m-1/2+\phi/\pi})(1+q^{m-1/2})^{2}
\right].
\end{eqnarray}
The supertrace now follows using expression (\ref{strace}). To get the idea why it is so, let us extract the contributions from the low-lying states corresponding to the scalars and the fermions in the hypermultiplet of the gauge theory:
\begin{equation}
\mathcal{Z}_{\mbox{{\scriptsize hyper}}} =q^{(\Delta
s)^2/\aprime-\phi/(2\pi)}+q^{(\Delta s)^2/\aprime+\phi/(2\pi)}
-2q^{(\Delta s)^2/\aprime}.
\end{equation}
Notice that the exponents of $q$ are indeed $\aprime M^2$ of the scalars and the fermions. Thus the supertrace formula (\ref{strace}) gives us
\begin{equation}
\left.{2 \over \aprime}q{\del \over \del q}
\mathcal{Z}_{\mbox{{\scriptsize hyper}}}\right|_{q=1}
=2\left((\Delta s)^2/(\aprime)^2-\phi/(2\pi\aprime)\right)
+2\left((\Delta s)^2/(\aprime)^2+\phi/(2\pi\aprime)\right)
-4(\Delta s)^2/(\aprime)^2.
\end{equation}
This is exactly the supertrace of $M^2$ for the ${\cal N}=2$ hypermultiplet.

Now it is straightforward to compute the supertrace (\ref{strace}). In terms of $\vartheta$-functions, $\vartheta_{\alpha\beta}(\nu,it)$, and the Dedekind $\eta$-function, $\eta(t)$, the partition function is given by

\begin{eqnarray}
\mathcal{Z}&=&\half e^{-2\pi t(\Delta s)^2/\aprime}\eta(t)^{-6}
\frac{\vartheta_{10}(0,it)\vartheta_{00}(0,it)}
{\vartheta_{01}(\phi t/(i\pi),it)\vartheta_{01}(0,it)}\nn\\
&&\qquad\times
\left[\vartheta_{00}(0,it)\vartheta_{10}(\phi t/(i\pi),it)
-\vartheta_{10}(0,it)\vartheta_{00}(\phi t/(i\pi),it)
\right].
\end{eqnarray}
Since we are interested in the behavior of the partition function at small $t$ or equivalently at $q\to1$, we will use the modular transformations for $\vartheta$-functions:
\begin{eqnarray}
\vartheta_{01}(\phi t/(i\pi),it)&=&
2\cos\phi e^{t\phi^2/\pi-\pi/(4t)}
t^{-1/2}\prod_{m=1}^{\infty}
(1-e^{-2\pi m/t})(1+e^{-2i\phi-2\pi m/t})
(1+e^{2i\phi-2\pi m/t}),\nn\\
\vartheta_{00}(\phi t/(i\pi),it)&=&
t^{-1/2}e^{t\phi^2/\pi}\prod_{m=1}^{\infty}
(1-e^{-2\pi m/t})
(1+e^{-2i\phi-2\pi m/t+\pi/t})
(1+e^{2i\phi-2\pi m/t+\pi/t}),\nn\\
\vartheta_{10}(\phi t/(i\pi),it)&=&
t^{-1/2}e^{t\phi^2/\pi}\prod_{m=1}^{\infty} (1-e^{-2\pi m/t})
(1-e^{-2i\phi-2\pi m/t+\pi/t}) (1-e^{2i\phi-2\pi m/t+\pi/t}),\nn\\
\eta(t)&=&t^{-1/2}e^{-\pi/(12t)} \prod_{m=1}^{\infty}(1-e^{-2\pi
m/t}). \label{modular}
\end{eqnarray}
Then one can find
\begin{equation}
{\del \over \del t}\mathcal{Z}
=0+{\cal O}(t)+{\cal O}(e^{-\pi/t}).
\end{equation}
Therefore the supertrace of $M^2$ is vanishing. In fact the
supertrace vanishes at each level, as we show explicitly in figure
5 for the first three levels.

\section {One-loop open string potential}
We will give some details of the one-loop computation of the $D4$+$D6$ system in this appendix. Starting from (\ref{bpot}) we have

\begin{eqnarray}
V &=&\int_0^{\infty} {dt \over t}\int_{-\infty}^{\infty}
\frac{d^4p}{(2\pi)^4} \exp\left[-2\pi t\left(\aprime p^2 +{(\Delta s)^2 \over
\aprime}\right)\right]
\nn \\
&\times&\left[- \Tr_{NS}\left(e^{2\pi t\left({\phi \over
2\pi}\right)} {1+(-1)^F \over 2} e^{-2\pi
t\left(N_B+N_F^{NS}+N_{bc}+N_{\beta\gamma}^{NS} \right)}
\right)\right. \\
&& \left. +\Tr_{R}\left({1+(-1)^F \over 2} e^{-2\pi
t\left(N_B+N_F^{R}+N_{bc}+N_{\beta\gamma}^{R} \right)}
\right)\right]\ . \nn
\end{eqnarray}
Using the expressions for the partition functions above, the one-loop amplitude amounts to
\begin{eqnarray}
V&=&\half\left({1 \over 8 \pi^2 \aprime}\right)^2
\int_0^{\infty} {dt \over t^3}
\exp\left[-2\pi t\left(
{(\Delta s)^2 \over \aprime}\right)\right]
{\eta(t)^{-6} \over \vartheta_{01}(0,it)
\vartheta_{01}({\phi t \over i\pi},it)} \nn \\
&\times&
\left[
\vartheta_{00}(0,it)^2\vartheta_{10}(0,it)
\vartheta_{10}({\phi t \over i\pi},it)
-\vartheta_{10}(0,it)^2\vartheta_{00}(0,it)
\vartheta_{00}({\phi t \over i\pi},it)
\right].
\end{eqnarray}
Note that when the angle $\phi$ is zero, the $D4$+$D6$ system is
supersymmetric and indeed the above one-loop amplitude is
vanishing, as a consequence of the vanishing of $\mathcal{Z}(q)$.

We are interested in the asymptotic behavior of the one-loop a
mplitude when the distance $\Delta s$ between $D4$ and $D6$-branes
is much larger than the string scale. This corresponds to the long
range force between two $D$-branes mediated by the exchange of
massless closed strings. When the separation $(\Delta
s)^2/\aprime$ is very large, only the region $t\to 0$ will
contribute to the one-loop amplitude $V$. By applying the modular
transformations of $\vartheta$-functions (\ref{modular}, one finds
the result shown in eq. (\ref{stringpot}).


\end{document}